\documentclass[aps,prd,nopacs,nofootinbib,notitlepage,twocolumn,longbibliography]{revtex4-1}

\usepackage{amsfonts}
\usepackage{amsmath}
\usepackage{mathtools}
\usepackage{xcolor}
\usepackage{bm}
\usepackage{fancybox}
\usepackage{comment}
\usepackage{multirow}
\usepackage{tipa}
\usepackage{longtable}
\definecolor{refs}{RGB}{245,156,74}
\usepackage[colorlinks=true,hyperfootnotes=true,citecolor=cyan]{hyperref}
\usepackage{enumitem}

\newcommand{\be}{\begin{equation}}
\newcommand{\ee}{\end{equation}}
\newcommand{\ba}{\begin{eqnarray}}
\newcommand{\ea}{\end{eqnarray}}
\newcommand{\bs}{\begin{subequations}}
\newcommand{\es}{\end{subequations}}

\newcommand{\bbe}{\boldsymbol{\mathrm{e}}}

\newcommand{\bbie}{\boldsymbol{\textbf{\textschwa}}}

\newcommand{\bkappa}{\boldsymbol{\kappa}}

\newcommand{\bomega}{\boldsymbol{\omega}}
\newcommand{\bTheta}{\boldsymbol{\Theta}}

\newcommand{\bA}{\boldsymbol{A}}

\newcommand{\bB}{\boldsymbol{B}}

\newcommand{\bE}{\boldsymbol{E}}
\newcommand{\bF}{\boldsymbol{F}}

\newcommand{\bh}{\boldsymbol{h}}
\newcommand{\bJ}{\boldsymbol{J}}
\newcommand{\bj}{\boldsymbol{j}}
\newcommand{\bL}{\boldsymbol{L}}
\newcommand{\bM}{\boldsymbol{M}}
\newcommand{\bO}{\boldsymbol{O}}

\newcommand{\bR}{\boldsymbol{R}}
\newcommand{\bT}{\boldsymbol{T}}
\newcommand{\bt}{\boldsymbol{t}}

\newcommand{\bX}{\boldsymbol{X}}

\newcommand{\bdiff}{\boldsymbol{\mathrm{d}}}
\newcommand{\bDiff}{\boldsymbol{\mathrm{D}}}

\newcommand{\lp}{\left(}
\newcommand{\rp}{\right)}

\newcommand{\nn}{\nonumber}

\newcommand{\1}{1$^\text{st}$}
\newcommand{\2}{2$^\text{nd}$}

\newcommand{\+}{ \prescript{+}{}}
\newcommand{\pox}{\boldsymbol{\Box}}

\newcommand{\TGZ}[1]{{\color{blue}\tt [TZ: #1]}}
\newcommand{\LM}[1]{{\color{red}\tt [LM: #1]}}
\newcommand{\PG}[1]{\textcolor{red}{PG: #1}}
\newcommand{\tomi}[1]{{\color{green}\tt [TK: #1]}}

\begin{document}

\title{Consistent first order action functional for gauge theories}

\date{\today}

\author{Priidik Gallagher}
\address{Laboratory of Theoretical Physics, Institute of Physics, University of Tartu, W. Ostwaldi 1, 50411 Tartu, Estonia}
\author{Tomi S. Koivisto}
\email{tomi.koivisto@ut.ee}
\address{Laboratory of Theoretical Physics, Institute of Physics, University of Tartu, W. Ostwaldi 1, 50411 Tartu, Estonia}
\address{National Institute of Chemical Physics and Biophysics, R\"avala pst. 10, 10143 Tallinn, Estonia}
\author{Luca Marzola}
\address{National Institute of Chemical Physics and Biophysics, R\"avala pst. 10, 10143 Tallinn, Estonia}
\author{Ludovic Varrin}
\address{National Centre for Nuclear Research, Pasteura 7, 02-093, Warsaw, Poland}
\address{Institute of Theoretical Physics and Astrophysics,
University of Gda\'{n}sk,
80-308 Gda\'{n}sk, Poland}
\author{Tom Zlosnik}
\address{Institute of Theoretical Physics and Astrophysics,
University of Gda\'{n}sk,
80-308 Gda\'{n}sk, Poland}

\begin{abstract}

A novel first order action principle has been proposed as the possible foundation for a more fundamental theory of 
General Relativity and the Standard Model. It is shown in this article that the proposal consistently incorporates
gravity and matter fields, and guides to a new and robust path towards unification of fundamental interactions.

\end{abstract}

\maketitle

\section{Introduction}

Lorentz symmetry is a cornerstone of modern physics. 
The Standard Model is formulated as a quantum field theory based on the global Lorentz symmetry of Special Relativity, the fields being classified according to the representations of the (complexified) Lorentz group \cite{Weinberg:1995mt}. Whilst gravity has been understood to 
arise from the ``gauging'' of the Poincar{\'e} group of the inhomogeneous Lorentz transformations in 
the Einstein-Cartan-Sciama-Kibble theory\footnote{This procedure consists of a combined gauging of the global ``internal'' Lorentz symmetry of fermionic actions and promotion of the symmetry of Standard Model actions under diffeomorphisms generated by the Killing vectors of Minkowski space (whose commutators satisfy the Lie algebra of the Poincar\'{e} group) to a full diffeomorphism symmetry. It could be argued that the latter part of the procedure is superfluous both mathematically (as manifest in the by-construction diffeomorphism-invariant language of differential forms) and physically (the introduction of the corresponding gauge force is not supported by the interpretation of gravity according to the equivalence principle) \cite{Koivisto:2019ejt}.}
and its generalisations \cite{Kibble:1961ba,hehl},
this has not yet lead to a reconciliation of General Relativity and quantum mechanics. 

A new take on the gauge theory of spacetime and gravity is based on precisely the homogeneous 
(complexified) Lorentz group\footnote{Possible formulations of a Lorentz gauge theory of gravity  
had been considered earlier \cite{Utiyama:1956sy,PhysRevLett.33.445,Aldrovandi:2004uz,Diakonov:2011im,Obukhov:2012je}, but a key point of the new theory \cite{Zlosnik:2018qvg} is the 
realisation that chiral asymmetry \cite{Plebanski:1977zz,Ashtekar:1986yd} is required for the existence of a General-Relativistic limit to the solutions
\cite{Nikjoo:2023flm}.} \cite{Zlosnik:2018qvg}. 
In general, gravitational models with polynomial actions can accommodate the zero ground state of the metric \cite{Witten:1988hc,Giddings:1991qi,Wilczek:1998ea,Banados:2007qz}, which we refer to as the ``pregeometric'' property \cite{Wheeler:1957mu,Akama:1978pg,Wetterich:2003wr,Wetterich:2021hru}.
%The Lorentz gauge theory is pregeometric \cite{Wheeler:1957mu,Akama:1978pg,Wetterich:2003wr,Wetterich:2021hru}, in the sense that it accommodates the zero ground state of geometry \cite{Witten:1988hc,Giddings:1991qi,Wilczek:1998ea,Banados:2007qz} %\TGZ{Possible to say a bit more about how this is defined and whether, say, Einstein-Cartan gravity allows it?}. \LM{ground state of what?} 
%The observer is rooted into the theory at a more fundamental level than in standard approaches, 
The natural idea that spacetime arises 
via a spontaneous symmetry breaking that selects a preferred direction of time \cite{Brown:1994py,Jacobson:2000xp,Gielen:2012fz} is often implemented by additional fields on top of the geometry, but in-built to the Lorentz gauge theory wherein the symmetry breaking is necessary to emerge from the pregeometric state. The subtle elaboration of the mechanism entails an apparently drastically different description of gravity and spacetime, where even the Minkowski space has dynamical curvature and torsion \cite{Koivisto:2022uvd}. A recent Hamiltonian analysis established the consistency of the Lorentz gauge theory \cite{Nikjoo:2023flm}, and the possibility of a new cosmological paradigm was speculated \cite{Koivisto:2023epd}.

In view of the SO(10) grand unification of the Standard Model gauge interactions \cite{Baez:2009dj}, the 
new SO$_{\mathbb{C}}$(1,3) $\cong$ SO$_{\mathbb{C}}$(4) gravitational gauge theory would naturally seem to fit into a yet grander SO(N) unification 
along the lines of the gravi-GUT proposals \cite{Percacci:1984ai,Nesti:2009kk,Chamseddine:2016pkx,Maiezza:2021dui,Konitopoulos:2023wst,Chkareuli:2023krd}. 
However, the coupling of the Standard Model to the Lorentz gauge theory calls for a pregeometrisation of also 
the internal gauge field sector \cite{Gallagher:2022kvv}. Whereas the standard spinor actions are polynomial in the
fields, \1 order in the derivatives and possess the pregeometric property, 
%these 3 desiderata  \TGZ{3 desiderata?} \LM{fancy word} 
%required a more fundamental action principle for the Yang-Mills gauge bosons. 
a more fundamental action principle was required for the Yang-Mills gauge bosons.
The suggested theory 
\cite{Gallagher:2022kvv} can differ already classically from previous \1 order 
formulations \cite{Halpern:1977fw,Okubo:1979gt,Kiriushcheva:2011aa,Lavrov:2021pqh}. 

The characteristic feature of the new \1 order gravity is 
the appearance of an effective dark matter source term. Interestingly, it was recently
pointed out by Kaplan {\it et al} \cite{Kaplan:2023wyw,Kaplan:2023fbl}, that since unitary evolution in quantum mechanics is described by
the Schr{\"o}dinger equation which is \1 order in time derivative, the classical limit of gauge theories, including 
gravity, could be generalised by the addition of {\it shadow charges}, whose presence reflects the fact that
quantum fluctuations need not satisfy the constraints imposed by the standard, \2 order
formulation of gauge interactions. This motivates us to consider also a modified version of the
\1 order Yang Mills theory, wherein shadow charges could arise as 
integration constants in the solutions to the equations of motion, analogously to the theory of 
gravity \cite{Gallagher:2021tgx,Gallagher:2022kvv}. 

We shall focus on the conserved charges in the framework of Lorentz gauge theory from the perspective
of the Noether's theorems, taking advantage of some recent developments in covariant phase space formalism \cite{Iyer:1994ys,Harlow:2019yfa,Freidel:2020xyx,Donnelly:2022kfs,delValleVaroGarcia:2022ceq,Gomes:2022vrc,Ciambelli:2022vot,Ciambelli:2023bmn}. 
This article reports the results of our derivations. Section \ref{actionprinciple} presents the action, Section \ref{sec:internal} covers the currents in the
gravitational sector and Section \ref{sources} covers the rest. All the are charges unambiguous and have a clear physical interpretation. We conclude in 
Section \ref{conclusion} that the consistency of the \1 action formulation provides a valuable guiding principle in the
quest for the final theory. 
%that the complete system of charges in the unified formulation of spacetime, matter and gauge fields needs no ``improvements'' and presents no ambiguities.

%of the Noether charges in the framework of Lorentz gauge theory. Though the results are rather simple,  
%the derivations involved many subtleties which we could only set straight
%thanks to quite recent, significant progress in the understanding of the  
%physical content of Noether's theorems
%None of the results were published previously.

\section{The action principle}
\label{actionprinciple}

We consider an action $I=\int\bL$ with the 4-form
\be \label{action}
\bL = \bL_{G} + \bL_M\,,
\ee
where $\bL_{G}$ is the gravitational Lagrangian four-form polynomial in the gravitational fields which are taken to be a connection for the (complexified) Lorentz group $\bomega^{ab}$ and a scalar field $\phi^{a}$ valued in the group's fundamental representation (which we term the \emph{khronon} due to its potential to introduce a standard of time into gravitation). We choose
\begin{align}
\bL_{G} &= \bB^{ab}\wedge\+\bR_{ab} \label{grav_lagrangian}
\end{align}
where we have introduced the short-hand for the (proto)area-element $\bB^{ab}$,
\be \label{proto}
\bB^{ab} = \frac{i}{2}{}\lp\bDiff\phi^a\wedge\bDiff\phi^b\rp\,,
\ee
and $\bR^{ab} = \bdiff\bomega^{ab}+ \ \bomega^{a}_{\phantom{a}c}\wedge \bomega^{cb}$ is the curvature 2-form for $\bomega^{ab}$. The ${}^\pm{}X=(1\mp i\star) X$ are the projectors to the self-dual (left-handed) or anti-selfdual (right-handed) sectors, $\star{}^{\pm}{}X = \pm i\, {}^{\pm}{}X$. 
It was demonstrated in \cite{Zlosnik:2018qvg} that (\ref{grav_lagrangian}) realizes an extension to General Relativity,
when the metric tensor $g$ is identified as $g=\bDiff\phi^a\otimes\bDiff\phi_a$. 

In (\ref{action}) we take into account minimally coupled matter fields $\psi$ which may be some $p$-forms.
$\bL_{M}= \bL_M(\bDiff\phi^a,\psi,\+\bDiff\psi)$ is the Lagrangian four-form for $\psi$ which includes the gravitational fields, but we have excluded non-minimal couplings of $\prescript{-}{}\bomega$ to $\psi$. We parameterise the material energy current $\bt_a$ and the spin current $\bO_{ab}$, respectively, as
\ba \label{sourcecurrents}
\bt_a  =  -\frac{\partial\bL_M}{\partial \bDiff\phi^a}\,, 
\quad
\bO_{ab}  =  -\text{rep}_{ab}\psi\wedge\frac{\partial\bL_M}{\partial\bDiff\psi}\,, 
\ea
%\TGZ{If we have some coupling to a spinor field containing ${}^{-}\omega$, in principle this will contain some information about the metric additional to the couplings to $\bDiff\phi^a$ which seems to complicate the above definition - it looks like we don't allow for such couplings in the spinor Lagrangians}
where $\text{rep}_{ab}$ represents the Lorentz generator for $\psi$. Detailed examples are considered in \ref{sources}.

The variation of the total action  
\be \label{variation}
\delta\bL = \delta\phi^a\bE_a + \delta\bomega^{ab}\wedge\bE_{ab}
+ \delta\psi\wedge\bE_\psi + \bdiff\bTheta\,,
\ee
then yields the equations of motion (EoMs) for the khronon, the gauge potential and the matter fields, respectively,
%\PG{Should $\bDiff\bB_{ab}\to{}^+\bDiff\bB_{ab}$?}
\bs
\ba
\bE_a & = & -\bDiff\lp i\bDiff\+\pox\phi_a -\bt_a\rp\,,  \\
\bE_{ab} & = & \+\bDiff\bB_{ab} - i\phi_{[a}\bDiff\+\pox\phi_{b]} + \phi_{[a}\bt_{b]} -\bO_{ab}\,, \\
\bE_\psi & = & \frac{\partial\bL_M}{\partial\psi} + (-1)^{p+1}\bDiff\frac{\partial\bL_M}{\partial \bDiff\psi}\,,
\ea
\es
and the symplectic potential
\be \label{symplectic}
\bTheta = \delta\phi^a\lp i\bDiff\+\pox\phi_a -\bt_a\rp + \+\delta\bomega^{ab}\wedge\bB_{ab} + \delta\psi\wedge\frac{\partial\bL_M}{\partial\bDiff\psi}\,, 
\ee
where $\pox =\bDiff\bDiff$ is the curvature 2-form operator and $\+\pox = \+\bDiff\+\bDiff$. %\TGZ{and ${}^{+}\Box = {}^{+}D{}^{+}D$?}. 
This shows that the action is stationary on-shell given Dirichlet boundary conditions for the variations of the gravitational and matter fields \cite{Harlow:2019yfa}. There are no boundary conditions\footnote{This depends crucially on the precise form of the action functional, and would not hold e.g. for the choice $\bL=(i/2)\pox\phi_a\wedge\+\pox\phi^a +\bL_M$ which is equivalent to (\ref{action}) up to a total derivative.} for the antiself-dual potential $\prescript{-}{}\delta\bomega^{ab}$. The EoM's $\bE_a$ and $\bE_{ab}$ imply that on-shell $\approx$ 
\bs
\ba
i\bDiff\+\pox\phi^a & \approx & \bt^a + \bM^a\,, \label{efe1} \\
\+\bDiff\bB^{ab} & \approx & \phi^{[a}\bM^{b]} + \bO^{ab}\,,
\ea
\es
where $\bM^a$ is a 3-form that satisfies $\bDiff\bM^a =0$.

\section{Symmetries}
\label{sec:internal}

%\TGZ{Is it worth briefly stating how we define a symmetry to be? i.e. Under a transformation of dynamical fields the action changes by at most a boundary term}

We consider transformations $\delta$ that act on the dynamical fields.
The transformation is a symmetry of $\bL$ if $\delta\bL=\bdiff \boldsymbol{\ell}$, and exact if $\boldsymbol{\ell}=0$. Besides the Lorentz and diffeomorphism symmetry, the action (\ref{action}) has a peculiar shift symmetry. Below we report the currents $\bJ$ corresponding to the 3 classes of symmetry transformations. Each current is manifestly conserved on-shell, $\bdiff\bJ \approx 0$.
For a gauge symmetry, the current is on-shell an exact form, $\bJ \approx \bdiff\bj$, where $\bj$ is called the Noether-Wald charge \cite{Iyer:1994ys,Harlow:2019yfa}. The charges are given as the integrated $\oint \bj$ of the Noether-Wald charge over a closed surface.

%\TGZ{Should we also then introduce charges and for each of our examples show what the charge is? (Integration of the Noether potential 2-form over a corner?)}

\subsection{Lorentz transformation}

Consider a Lorentz transformation of the fields with  infinitesimal parameters $\lambda^a{}_b$,
\bs
\ba
\delta_\lambda\phi^a & = & \lambda^a{}_b\phi^b\,, \\
\delta_\lambda\bomega^a{}_b & = & -\bDiff\lambda^a{}_b\,, \\ 
\delta_\lambda \psi & = & \lambda^{ab}\text{rep}_{ab}\psi\,. 
\ea
\es
The Lorentz symmetry is exact $\delta_\lambda\bL=0$, and we
take this to be the case also independently for the matter 4-form $\delta_\lambda\bL_M=0$. Then we obtain Noether identities independently for the gravitational and matter sector. These are derived from (\ref{variation}) by considering parameters $\lambda^{ab}$ which vanish at the boundary s.t. we can neglect all the total derivatives in the variations. We obtain the 2 identities,
\bs
\ba
\+\pox\bB_{ab} & = & i\bDiff\phi_{[a}\wedge\bDiff\+\pox\phi_{b]}\,, \\ 
\bDiff\bO_{ab} & = & \bDiff\phi_{[a}\wedge\bt_{b]} + \text{rep}_{ab}\psi\wedge\bE_\psi\,.
\ea
\es
%For an exact symmetry, the Noether current is the symplectic potential generated by the transformation \TGZ{Maybe change wording - it's the symplectic potential inner-producted with a vector field on field configuration space}
The Noether current
\be
\bJ_\lambda = \lambda^{ab}\bE_{ab} - \bdiff\lp \+\lambda^{ab}\bB_{ab}\rp\,, \label{lcurrent}
\ee 
%\TGZ{Worth mentioning somewhere that all these $\textbf{J}$ have $dJ=0$ when the equations of motion are satisfied?}
is an exact form on-shell $\bJ_\lambda \approx \bdiff\bj_\lambda$, where the Noether charge 2-form is now $\bj_\lambda=\+\lambda^{ab}\bB_{ab}$. Only the self-dual Lorentz transformations are associated with non-trivial charges.  

\subsection{Shift symmetry}

The action $\int\bL$ enjoys a shift symmetry, the invariance under constant translations of the khronon\footnote{Perchance this could be understood due to $\phi^a$ representing the symmetry of not the group but the torsor https://math.ucr.edu/home/baez/torsors.html.},
\bs
\ba
\delta_\chi\phi^a & = & \chi^a \quad \text{where} \quad \bDiff\chi^a =0\,, \\
\delta_\chi\bomega^a{}_b & = & 0\,, \quad \delta_\chi \psi  = 0\,. 
\ea
\es
The Noether identity is trivial for this transformation. The charge that we obtain using (\ref{symplectic}) and then (\ref{efe1})
\be
\bJ_\chi  =  \chi^a\lp i\bDiff\+\pox\phi_a -\bt_a\rp 
\approx  \chi^a\bM_a\,, 
\ee
describes the energy-momentum carried by the effective matter 3-form $\bM_a$. This can be contrasted with Poincar{\'e} gauge theory, where the local translation is called a trivial gauge symmetry since it has zero charge. (One has to break covariance in order to extract a nonzero charge. We'll return to this point at \ref{noncovariance}.) 

%The Noether potential is $\bj_\chi=-\xi^a\bE_a$. 

\subsection{Diffeomorphism}

In the Lorentz gauge theory, spacetime geometry (coframe and curvature) is generated by Lie-dragging the fundamental fields (khronon and gauge potential)  \emph{covariantly}\footnote{The transformation can be considered as the minimal coupling of the frame-dependent definition discussed below \ref{noncovariance}.} along a vector $\xi$:
\bs
\ba
\delta_\xi \phi^a & = & \xi\lrcorner\bDiff\phi^a\,, \\
\delta_\xi \bomega^a{}_b & = & \xi\lrcorner\bR^a{}_b\,, \\
\delta_\xi \psi & = & \{\xi\lrcorner,\bDiff\}\psi\,.
\ea
\es
where $\lrcorner$ is the interior product on differential forms, and here and in what follows, the $\bDiff$ is always the total covariant derivative, thus involving also internal gauge fields in the case that the fields $\psi$ have internal gauge charge.
This gauge symmetry is not exact in the sense of $\bL$ being invariant under the transformation, but $\delta_\xi\bL = \bdiff(\xi\lrcorner\bL)$. We obtain the 
Noether identity for gravity, 
\bs
\be
i(\xi\lrcorner\bDiff\phi^a)\pox\+\pox\phi_a = \xi\lrcorner\bR^{ab}\wedge\lp \+\bDiff\bB_{ab} -i\phi_{[a}\bDiff\+\pox\phi_{b]}\rp\,,\nn
\ee
and for the invariance of $\int\bL_M$ we get 
\be
(\xi\lrcorner\bDiff\phi^a)\bDiff\bt_a + \xi\lrcorner\bR^{ab}\wedge\lp\phi_{[a}\bt_{b]}-\bO_{ab}\rp = -\delta_\xi \psi\wedge\bE_\psi\,.\nn
\ee
\es
In a non-degenerate spacetime wherein $\bbe^a\equiv\bDiff\phi^a$ has an inverse $\bbie_a$, these can be rewritten as 
\bs
\begin{align}
i\pox\+\pox\phi_a  & =   \bbie_a\lrcorner\bR^{bc}\wedge\+\bDiff\bB_{ab} + i\bT^a\wedge\bDiff\+\pox\phi_a\,,\,\, \\
 -\delta_{\bbie_a} \psi\wedge\bE_\psi & =  \bDiff\bt_a -\bbie_a\lrcorner\bT^b\wedge\bt_b - \bbie_a\lrcorner\bR^{bc}\wedge\bO_{bc}\,, \quad \,.
\end{align}
\es
where $\bT^a=\bDiff\bbe^a=\pox\phi^a$. 
The Noether current vanishes identically $\bJ_\xi = \xi\cdot\bTheta - \xi\lrcorner\bL=0$, and thus implies that a change of coordinates is a trivial gauge transformation. %\TGZ{Worth comparing to standard result (?) in Minkowksi space that symmetry of matter actions in fixed Minkowski background under diffeomorphisms corresponding to the Killing vectors of Minkowski space can - with 'improvements' - lead to the conservation of the stress energy momentum tensor and the six conservations associated with the boost and rotation Killing vectors?}. 
The matter sources have to be formulated consistently s.t.
\be \label{hilbert}
(\xi\lrcorner\bDiff\phi^a)\bt_a = \delta_\xi\psi\wedge\frac{\partial \bL_M}{\partial \bDiff\psi} - \xi\lrcorner\bL_M\,,
\ee
which means that the Hilbert (i.e. the metrical) and the Noether (i.e. the canonical) energy-momenta are equivalent.%\TGZ{Either a citation here or a sentence/footnote defining the two different energy momenta?}.  

\subsection{On frame-dependent charges}
\label{noncovariance}

One can combine transformations from the above 3 classes of symmetry transformations. 
An example is the coordinate diffeomorphism,
\bs
\label{cdiff}
\ba
\mathcal{L}_\xi \phi^a & = & \xi\lrcorner\bdiff\phi^a\,, \\
\mathcal{L}_\xi \bomega^{ab} & = & \bDiff\lp\xi\lrcorner\bomega^{ab}\rp + \xi\lrcorner\bR^{ab}\,, \\
\mathcal{L}_\xi \psi & = & \{\xi\lrcorner,\bdiff\}\psi\,,
\ea
\es
which is the combination of a Lorentz transformation and a proper diffeomorphism,
$\mathcal{L}_\xi = \delta_\xi + \delta_{\lambda=\xi\lrcorner\bomega}$. %\TGZ{It might be useful to write this around eqs 14a just to clarify to the reader how these covariant diffs are constructed from the non-covariant ones + Lorentz transformations?}. 
The possible physical
relevance of this transformation is subject to case-dependent subtleties. The way that the fields are dragged along a vector $\xi$ has no Lorentz-covariant meaning. The corresponding charge has no Lorentz-invariant interpretation. With some manipulations, using e.g. (\ref{sourcecurrents}) and assuming (\ref{hilbert}), one can verify that the Noether current from (\ref{cdiff}) is given, as expected, precisely by (\ref{lcurrent}) with the Lorentz transformation parameter $\lambda^{ab}=\xi\lrcorner\bomega^{ab}$. So, the charge is frame-dependent because the parameter is non-covariant. 

Nevertheless, it is very well known that the currents generated by $\mathcal{L}_\xi$ correctly describe the physical energy and momenta in many relevant special cases.
This is so because energy and momentum can only be defined with respect to a reference frame, and thus it is expected that these charges are frame-dependent\footnote{According to a recent proposal,
the frame-dependence is the consequence of the equivalence principle, and the physical criterion that uniquely fixes the reference frame is the vanishing of its local energy-momentum current \cite{Gomes:2022vrc}. However, it is outside this article's scope to implement this so called G$_\parallel$R principle \cite{Gomes:2023hyk} in the Lorentz gauge theory.}. The basic example is the standard result in Minkowski space that the symmetry of matter actions in the fixed background under diffeomorphisms corresponding to the Killing vectors of Minkowski space can - with ``improvements'' - lead to the conservation of the stress energy momentum tensor and the six conservations associated with the boost and rotation Killing vectors. This can be generalised to a maximally symmetric space, available perhaps globally, locally, asymptotically, or say, as an extra-dimensional embedding. These considerations apply as such in the geometric phase of Lorentz gauge theory. 

%\LM{This thing really confuses me because Nother charges should be constants of motion. How can these be frame-dependent? }\TGZ{Can the Wald 'Black hole entropy is Noether charge' stuff still be recovered in the khronon model?}

\section{Sources}
\label{sources}

We consider fermions below in \ref{fermions}, gauge bosons in \ref{bosons}, and scalars in \ref{scalars}. A unimodular version of the theory
is briefly checked in \ref{uni}.

\subsection{Fermion matter}
\label{fermions}

%\TGZ{Possible change away from $\psi$ for Fermion fields as we earlier use $\psi$ to mean a generic matter field?}
Dirac's theory of the electron and Weyl's theory of the neutrino pass 
the pregeometric standards and need no modifications.
%Though in section \ref{actionprinciple} we denoted generic matter sources by $\psi$, let $\psi$ in this subsection denote spin-$1/2$ fields\PG{Maybe just "Let here/in this subsection $\psi$ denote a dirac spinor." Or some similar rephrasing.}.
Let $\psi$ in this subsection denote the Dirac spinor. 
The $\gamma_a$ in the spin-$1/2$ $\text{rep}_{ab} = -\gamma_{[a}\gamma_{b]}/2$ are matrices which obey $\gamma_{(a}\gamma_{b)} = -\eta_{ab}$. The Dirac spinor $\psi$ has the conjugate $\bar{\psi}=\psi^\dagger\gamma^0$. In this representation, 
$\star = i\gamma^5 = -\gamma^0\gamma^1\gamma^2\gamma^3$, and we can project the 2 Weyl spinors $\prescript{\pm}{}\psi = (1\mp\gamma^5)\psi/2$. %\LM{This convention may be confusing; I would use ${}^\pm\psi$.} 
Define also $\star\bDiff\phi^a = \epsilon^a{}_{bcd}\bDiff\phi^b\wedge\bDiff\phi^c\wedge\bDiff\phi^d/3!$ and $\star 1 = \epsilon_{abcd}\bDiff\phi^a\wedge\bDiff\phi^b\wedge\bDiff\phi^c\wedge\bDiff\phi^d/4!$.
Then, adopting the prescription of Ref.\cite{Ashtekar:1989ju},
\be
\bL_M = \frac{i}{2}\lp\star\bDiff\phi^a\rp\wedge\lp \bar{\psi}\gamma_a\bDiff\+\psi - \bDiff\bar{\psi}\gamma_a\prescript{-}{}{\psi}\rp 
- \bar{\psi}\psi\star m\,.
\ee
%\TGZ{!Define notation for the definition of $\star$ and $\star\bDiff\phi^{a}$}
From the variation
\ba
\delta\bL_M & = &  -\delta\lp\bDiff\phi^a\rp\wedge\bt_a + \delta\bomega_{ab}\wedge\delta\bO^{ab} \nn \\
& + & \delta\bar{\psi}\bE_{\bar{\psi}} + \bE_{\psi}\delta\psi + \bdiff\bTheta\,, 
\ea
we obtain the currents
\bs
\ba
\bt_a & = & \lp\star\bB_{ab}\rp\wedge\lp \bar{\psi}\gamma^b\bDiff\+\psi - \bDiff\bar{\psi}\gamma^b\prescript{-}{}{\psi}\rp 
\nn \\
& + & m\bar{\psi}\psi\star\bDiff\phi_a\,,\quad \label{fenergy} \\
\bO^{ab} & = & \frac{i}{8}\lp\star\bDiff\phi_c\rp\bar{\psi}\lp \gamma^c\gamma^{[a}\gamma^{b]}\+\psi  + \gamma^{[a}\gamma^{b]}\gamma^c\prescript{-}{}\psi\rp \nn \\
& = & \frac{i}{2}\bar{\psi}\+{}\lp\star\bDiff\phi^{[a}\gamma^{b]}\rp\gamma^5\psi\,,
\ea 
\es
the EoMs
\bs
\ba
\bE_{\bar{\psi}} & = & \frac{i}{2}\gamma_a\lp \star\bDiff\phi^a\rp\wedge\bDiff\psi - \gamma^a\lp\star\bB_{ab}\rp\wedge\bT^b\prescript{-}{}\psi  \nn \\
& - & \psi\star m\,, \\ 
\bE_{{\psi}} & = & -\frac{i}{2}\lp\star\bDiff\phi^a\rp\wedge\bDiff\bar{\psi}\gamma_a
+ \lp\star\bB_{ab}\rp\wedge\bT^b\+{}\bar{\psi}\gamma^a \nn \\
& - & \star m \bar{\psi}\,, 
\ea
\es
and the symplectic potential
\be
\bTheta = \frac{i}{2}\lp\star\bDiff\phi^a\rp\lp \delta\bar{\psi}\gamma_a\prescript{-}{}\psi -
\bar{\psi}\gamma_a\delta\+\psi\rp\,. 
\ee
In a real frame, $\bE_{\bar{\psi}}=\bar{\bE}_\psi$. The identity (\ref{hilbert}) is consistent with the energy current (\ref{fenergy}). 

\subsection{Yang-Mills fields}
\label{bosons}

The \1 order pregeometric Yang-Mills theory \cite{Gallagher:2022kvv} is formulated in terms of the interface (proto)area element
\be \label{proto2}
\tilde{\bB}{}^{ab} = \bh^a\wedge\bDiff\phi^b\,,
\ee
with the ``one foot outside'' and the other $\bh^a$, valued in the adjoint representation of the Yang-Mills gauge group,  a ``vierbein'' spanning an internal hyperspace\footnote{On frames constructed from material fields in condensed matter physics, see \cite{Volovik:2023faj}.}. We recall that $\bDiff$ is the total covariant derivative, thus involving also the Yang-Mills gauge field $\bA$ whose field strength is denoted by $\bF$.  Now the field excitation $\ast\bF$ (where $\ast$ is the Hodge dual) is not postulated a priori, but the gist of this new approach to gauge interactions is that the field excitation  $\ast\tilde{\bB}=\eta^{ab}\ast\tilde{\bB}_{ab} \approx \ast \bF$ emerges from the variational principle. An action density which achieves this is
\be \label{YMaction}
\bL_M = \langle \tilde{\bB}{}^{ab}\wedge( \star\tilde{\bB}{}_{ab} - \eta_{ab}\bF )\rangle\ - \langle\bA\wedge\tilde{\bJ}\rangle\,,
\ee
%\TGZ{We might have to clarify that $\mathbf{DA}$ is the curvature and not the covariant derivative applied to a field in the adjoint representation?}
%\TGZ{Do we recover the gravity action when we set $\alpha^{a,bc}=\epsilon^{abcd}\phi_{d}$ and $\mathbf{A} = \mathbf{\omega}^{+}$? If seems there would be the cosmological constant term, and the gravity term?}
where $\bA$ is the Yang-Mills gauge field, $\tilde{\bJ}$ is its material source, and $\langle \cdot \rangle$ is the trace over the Lie algebra.

\subsubsection{Standard theory}

The variation
\ba \label{YMvariation}
\delta\bL_M & = & -\delta\lp \bDiff\phi^a\rp\wedge\bt_a + \delta\bomega^{ab}\wedge\bO_{ab} \nn \\ & + & \langle\delta\bh^a\wedge\tilde{\bE}{}_a\rangle 
+ \langle\delta\bA\wedge\tilde{\bE}\rangle + \bdiff\bTheta\,,  
\ea
yields us the EoMs,
\bs
\label{YMeom}
\ba
\tilde{\bE}{}_a & = & -2\star\tilde{\bB}{}_{ab}\wedge\bDiff\phi^b + \bF\wedge\bDiff\phi_a\,,  \label{YMeom1} \\
\tilde{\bE} & = & \bDiff\tilde{\bB} - \tilde{\bJ}\,,
\ea
\es
and the symplectic potential
\be \label{YMsymplectic}
\bTheta = 
- \langle\delta\bA\wedge\tilde{\bB}\rangle\,. 
\ee
The gravitational source currents are
\bs
\ba
\bt_a & = & 2\langle\star\tilde{\bB}_{ab}\wedge\bh^b\rangle - \langle\bF\wedge\bh_a\rangle\,, \label{YMt0}\\
\bO_{ab} & = & 0\,.\,\,\,
\ea
\es
It is not difficult to see that the internal symmetry transformation 
\ba
\delta_g \bh^a =  [g,\bh^a]\,, \quad
\delta_g \bA  =  -\bDiff g\,,
\ea
results in the expected current $\bJ_g \approx \tilde{\bJ}$. It has to be concluded that this prescription is the mere reformulation of the standard Yang-Mills theory. In particular, the symplectic current (\ref{YMsymplectic}) assumes its expected form, and the energy current (\ref{YMt0}) fails the consistency requirement (\ref{hilbert}). 

A slightly more economic reformulation considers instead the 6 d.o.f.'s of the excitation carried in the fundamental variational d.o.f. $\alpha^{ab}$ valued in the adjoints of both the Lorentz and the Yang-Mills gauge groups, s.t. $\bh^a=\alpha^a{}_b\bbe^b$. However, this would not change the conclusions. 
%Therefore, we are motivated to consider the somewhat more radical alternative, 

\subsubsection{Modified theory}

A more radical alternative is to encode the variational d.o.f.'s into the isokhronon $\alpha^a$ living in the fundamental representation of the Lorentz group and giving rise to the internal hyperspacetime $\bh^a = \bDiff\alpha^a$ in an analogy to the khronon $\phi^a$ in the external spacetime. Then an analogy of dark matter may also arise in form of nontrivial vacua. 
This describes the situation in quantum mechanics wherein the field force lines need not be strictly attached to the material source points. The case $\ast\tilde{\bB} \approx \ast\bF$ is just one of the solutions, and therefore the solution space can be constrained by phenomenological data\footnote{In cosmology \cite{Koivisto:2023epd} it remains to be seen whether $\bM_a$ could be related to dark matter and the $\bX_a$ in the result (\ref{theX}) of this section to magnetic fields.}.

%An action density which achieves this is\footnote{{The term involving $\bF$ is - up to a boundary term - $\langle\alpha_{a}\phi_{b}\bR^{ab}\!\wedge \bF\rangle$. The Lorentz gauge theory possesses solutions with Minkowski metric but with non-vanishing anti-self dual curvature ${}^{-}\bR^{ab}$ \cite{Koivisto:2022uvd} and it is on these backgrounds that the equations of motion determine $\alpha^{a}$ to depend on $\bF$ so as to reproduce familiar Yang-Mills dynamics.}}
%\be \label{YMaction}
%\bL_M = \langle \tilde{\bB}{}^{ab}\wedge( \star\tilde{\bB}{}_{ab} - \eta_{ab}\bF )\rangle\ - \langle\bA\wedge\tilde{\bJ}\rangle\,,
%\ee
%\TGZ{We might have to clarify that $\mathbf{DA}$ is the curvature and not the covariant derivative applied to a field in the adjoint representation?}
%\TGZ{Do we recover the gravity action when we set $\alpha^{a,bc}=\epsilon^{abcd}\phi_{d}$ and $\mathbf{A} = \mathbf{\omega}^{+}$? If seems there would be the cosmological constant term, and the gravity term?}
%where $\bA$ is the Yang-Mills gauge field, $\tilde{\bJ}$ is its material source, and $\langle \cdot \rangle$ is the trace over the Lie algebra. 
The variation (\ref{YMvariation}) should then be reconsidered,
\ba
\delta\bL_M & = & -\delta\lp \bDiff\phi^a\rp\wedge\bt_a + \delta\bomega^{ab}\wedge\bO_{ab} \nn \\ & + & \langle\delta\alpha^a\bDiff\tilde{\bE}{}_a\rangle 
+ \langle\delta\bA\wedge\tilde{\bE}\rangle + \bdiff\bTheta\,,  
\ea
since now the 3-form $\tilde{\bE}_a$ in (\ref{YMeom1}) is closed but may not vanish on-shell. Nontrivial modifications now enter into the expression for the
%\bs
%\label{YMeom}
%\ba
%\tilde{\bE}{}_a & = & -\bDiff\lp 2\star\tilde{\bB}{}_{ab}\wedge\bDiff\phi^b - \bF\wedge\bDiff\phi_a\rp\,, \\
%\tilde{\bE} & = & \bDiff\tilde{\bB} - \tilde{\bJ}\,,
%\ea
%\es
symplectic potential,
\be
\bTheta = \langle \delta\alpha^a( 2\star\tilde{\bB}_{ab}
- \eta_{ab}\bF )\rangle\wedge\bDiff\phi^b
- \langle\delta\bA\wedge\tilde{\bB}\rangle\,,
\ee
as well as the gravitational source currents,
\bs
\ba
\bt_a & = & 2\langle\star\tilde{\bB}_{ab}\wedge\bDiff\alpha^b\rangle - \langle\bF\wedge\bDiff\alpha_a\rangle\,, \label{YMt}\\
\bO_{ab} & = & \langle\alpha_{[a}\bDiff\phi_{b]}\wedge\bF\rangle
- 2\langle \alpha_{[a}\star\tilde{\bB}_{b]c}\rangle\wedge\bDiff\phi^c\,.\,\,\,
\ea
\es
Remarkably, the energy current (\ref{YMt}) identically satisfies (\ref{hilbert}). So, the results for the 3 classes of gravitational charges in \ref{sec:internal} remain intact in the presence of the modified Yang-Mills interactions. 

It can be verified that the internal symmetry transformation $\delta_g \alpha^a =  [g,\alpha^a]$, 
$\delta_g \bA  =  -\bDiff g$ is associated with the current 
%The Yang-Mills theory possesses also internal symmetries. The infinitesimal gauge transformation,
%\bs
%\ba
%\delta_g \alpha^a =  [g,\alpha^a]\,, \quad
%\delta_g \bA  =  -\bDiff g\,,
%\ea
%\es
%results in 
\ba \label{YMcurrent}
\bJ_g & = & -\langle g\tilde{\bE}\rangle + 
\bdiff\langle g \tilde{\bB}\rangle \nn \\
& + & \langle [g,\alpha^a]\lp 2\star \tilde{\bB}_{ab}-\eta_{ab}\bF\rp\rangle\wedge\bDiff\phi^b
\approx \tilde{\bJ}\,,\quad\,\,\, 
\ea
where in the last step we used the EoM's (\ref{YMeom}) (see III.C of \cite{Gallagher:2022kvv}). The possible contribution to the divergence of $\tilde{\bB}$ due to a vacuum polarisation or magnetisation (see Eq.(50) of \cite{Gallagher:2022kvv})) is cancelled by the \2 term in (\ref{YMcurrent}), and we recover the canonical gauge current. A novel property of isokhronon theory is the shift symmetry,
\be
\delta_{\tilde{\chi}} \alpha^a = \tilde{\chi}^a\,, \quad \text{where} \quad \bDiff \tilde{\chi}^a = 0\,. 
\ee
The conserved current,
\be
\bJ_{\tilde{\chi}} = \tilde{\chi}^a\lp 2\star\tilde{\bB}_{ab} - \eta_{ab}\bF\rp\wedge\bDiff\phi^b
\approx  \tilde{\chi}^a\bX_a\,, \label{theX}
\ee
is the integration form $\bX_a$ responsible for the possible vacuum excitation \cite{Gallagher:2022kvv}. It is the analogy of the integration form $\bM_a$ in the gravity sector\footnote{Indeed, we recover the gravity action with a cosmological constant when we set $\alpha^{a,bc}=\epsilon^{abcd}\phi_{d}$ and identify $\bA$ with $\+\bomega$. In this sense, the actions for Yang-Mills fields and gravitation have a similar character. A perturbative hint of this similarity is already well-known from the context of amplitudes, as the so called double copy structure \cite{Bern:2010ue}, manifest in (\ref{proto}) {\it vs} (\ref{proto2}).}.

An important caveat is that one is now not free to choose both integration forms independently for arbitrary solutions. Therefore this theory is probably not a viable modification of the Standard Model gauge interactions. Let us briefly speculate on a possible refinement of the unified theory, \1 restricting to case of an Abelian gauge field $\bA$. Now, if we consider, instead of $\phi^a$, a field in $((\frac{1}{2}\otimes\bar{0})_-\otimes(\frac{1}{2}\otimes\bar{\frac{1}{2}})_+)$ of the complex Lorentz group, and instead of the $\alpha^a$, a field in $((0\otimes\bar{\frac{1}{2}})_-\otimes(\frac{1}{2}\otimes\bar{\frac{1}{2}})_+)$, then both of these fields are coupled to an independent SU(2) connection. Consequently, there always exist solutions with $\bX_a=0$, apparently restoring the viable limit to standard gauge theory. However, this prescription is not without other repercussions as then the $\tilde{\bB}$ is not a scalar but carries the SU(2)$\times$SU(2) charges from the anti-selfdual sector of the Lorentz group. Optimistically, this hints to the structure of the gravielectroweak theory and to the geometrisation of the Higgs mechanism operated by the isokhronon in the hyperspacetime.  

\subsection{Scalar fields}
\label{scalars}
%\PG{Edit or remove for best purpose. Particularly, check the factorial contributions.}
%Pregeometric scalar field theory mimics the Yang-Mills fields both in approach and form, but non-interacting 0-form scalar functions $\zeta$ lack any Lie group or algebra relation\PG{If already referring using $\bDiff$, then this sentence should be rewritten. Decide what's the main result in this section!}, and the gradient $\bDiff\zeta$, being the closest analogue of field strength, is just a 1-form. Thus the auxiliary field needs to build a 3-form for a dual, with the simplest option being 0-form $G^{abc}$ and the first order Lagrangian
%\begin{equation}
%    \bL_M = G_{abc}\Sigma^{abc}\wedge\bDiff\zeta + \bigg(\frac{1}{4}G_{abc}G^{abc}+U(\zeta)\bigg)\star1 ,
%\end{equation}
%\TGZ{Define $\Sigma^{abc}$ and $\star 1$}
% 
Putting the above speculation aside, since the Standard Model features a Higgs scalar field, for completeness we take into account a scalar field $\zeta$. In the \1 order formulation, it is accompanied by a Lorentz vector $z_a$, and a possible action is
\be
\bL_M = z_a\star\bDiff\phi^a\wedge\bDiff\zeta + \lp\frac{1}{4}z_a z^a + U(\zeta)\rp\star 1\,,
\ee
leaving open the possibility of a nontrivial potential $U(\zeta)$.
%The analysis proceeds through the first variation, and, as before, the two equations of motion
%\begin{subequations}
%    \begin{align}
%        \bE_\zeta &= \bDiff(G_{abc}\Sigma^{abc}) + U'(\zeta)\star1,\\
%        \bE_G^{abc} &= \Sigma^{abc}\wedge\bDiff\zeta + \frac{1}{2}G^{abc}\star1,
%    \end{align}
%\end{subequations}
%provide the prototype wave equation and the duality constraint. 
We obtain the EoM
\bs
\ba
\bE_\zeta & = & \bDiff\lp z_a\star\bDiff\phi^a\rp + U'(\zeta)\star 1\,, \\
\bE^a & = & \star\bDiff\phi^a\wedge\bDiff\zeta + \frac{1}{2}z^a\star 1\,,
\ea
\es
the symplectic contribution
\be
\bTheta = - z_a\star\bDiff\phi^a\delta\zeta\,,
\ee
and the source current
\bs
\ba
\bt_a  =  i\epsilon_{abcd} z^b\bB^{cd}\wedge\bDiff\zeta - \lp \frac{1}{4}z_b z^b + U(\zeta)\rp\star \bDiff\phi_a\,, \quad 
%\bO_{ab} & = & 0\,.
\ea
\es
whilst for scalar fields $\bO_{ab}  = 0$.
%\TGZ{Second term in 37a some kind of index stuff going on?}\PG{$\star1$ should probably be $\star\bDiff\phi_a$.}
%The energy-momentum and hypermomentum contributions are default, with $\bDiff\zeta$ providing an addition to the symplectic potential by
%\begin{equation}
%    \bTheta = -\delta\zeta G_{abc}\Sigma^{abc}.
%\end{equation}
%Note the theory doesn't a priori contain any new, nontrivial symmetries outside the inherent Lorentz and diffeomorphism invariance, as the scalar shift $\zeta\to\zeta+\mathrm{const}$ can be spoiled by a nontrivial potential $U(\zeta)$, and the auxiliary $G^{abc}$ does not couple outside Lorentz contractions. However, the purpose of the known Higgs scalar field would be to instead break the electroweak symmetry of the Standard Model, which a first order theory is likewise fit to provide, but cannot be studied in isolation from other sectors.

%\PG{Check numerical coefficients, I currently just erased the factorials. It is probably better to define a simple 0-form $\bar{G}^a=\epsilon^a{}_{ijk}G^{ijk}$, which contracts with $\bar{G}_a\star\bDiff\phi^a$. Then everything is a vector.}

\subsection{Cosmological constant}
\label{uni}

The perhaps simplest energy source is a cosmological constant. The contribution to the matter action is given by a Lagrangian with 2 new fields, a scalar $\Lambda$ and a 3-form $\bkappa$,
\be \label{lambda}
\bL_M = \frac{1}{2}\Lambda\lp \bdiff\bkappa - \star 1\rp\,.
\ee
The source contributions (\ref{sourcecurrents}) is
\be
\bt^a = -\frac{1}{2}\Lambda\star\bDiff\phi^a\,, \quad \bO^{ab} = 0\,.
\ee
The EoM's for the 2 fields dictate that $\bdiff\bkappa \approx \star 1$ and $\bdiff\Lambda \approx 0$. Thus $\bL_M \approx 0$. In the derivation of the diffeomorphism Noether current, we have to take into account that now (\ref{hilbert}) does not hold. We obtain $\bJ_\xi = \Lambda \ast \xi/2$, so it would seem that the $\Lambda$ does contribute. The non-trivial charge reflects the effective breaking of the longitudinal diffeomorphisms.

The 3-form gauge symmetry $\bkappa \rightarrow \bkappa + \boldsymbol{k}$, where $\boldsymbol{k}$ is an arbitrary 2-form, has a non-trivial charge that is given as the integral of $\bj_{\bkappa} = \Lambda \boldsymbol{k}/2$ over a 2-surface.

\section{Conclusion}
\label{conclusion}

Conserved charges lie at the heart of gauge theories. They characterise the observables of the theory and their algebra governs the structure of the theory. Charges are of paramount importance in holography and play a central role in (most approaches to) quantum gravity. In fact, the putative quantum theory might be entirely deduced from the charge algebra, according to the corner proposal and related current developments \cite{Freidel:2020xyx,Donnelly:2022kfs}.

In this article we presented the physical charges in the new Lorentz gauge theory of spacetime and gravitation. The charges associated with the Lorentz symmetry and diffeomorphism symmetry are the direct extrapolation ($\bbe^a \rightarrow \bDiff\phi^a$) of the results in Poincar{\'e} gauge theory. A novel feature is the ``dark shadow matter'' current $\bM_a$ associated with the shift symmetry of the action (\ref{action}). 

The theory was coupled to the pregeometrised Standard Model of particle physics, and it was shown that its matter fields generate consistently both the energy momenta and the angular momenta source currents.
%and it was shown that both the matter fields and the gauge fields consistently generate both energy momenta and angular momenta source currents. 
%The \1 order Yang-Mills theory features the newly suggested ``shadow charges'', which are now associated with the shift symmetry of the fields $\alpha^a$ in the internal sector. 
However, the most straightforward implementation of the Standard Model gauge fields inherits the issue in their usual, \2 order geometric formulation, which does not consistently describe the gravitational sources by the canonical Noether currents. It has often puzzled theoreticians 
%since Noether's and Bessel-Hagen's days 
that the canonical energy-momentum currents have the wrong expression, unless modified by some of the proposed ``improvements'' \cite{Weinberg:1995mt,Gomes:2022vrc,BELINFANTE1939887,Callan:1970ze,Blaschke:2016ohs,Baker:2021hly,Kourkoulou:2022ajr}. 
We considered a possible modification of the pregeometric \1 order theory, which would provide a solution to the issue, and features the newly suggested shadow charges, associated with the shift symmetry of the \1 order fields in the internal sector. 

The modified theory is not yet a phenomenologically viable replacement of the Standard Model interactions (though it might describe hypothetical new interactions e.g. in cosmology), but calls for the elaboration towards a more final theory. We conclude that the \1 order action principle provides a new 
robust framework to negotiate the unification of internal and spacetime gauge interactions and the reconciliation of gravity and quantum mechanics. 

%That such ad hoc improvements are now avoided, and that the ``shadow-charged'' dynamical system arises consistently, is new and uniquely compelling evidence that 
%the Lorentz gauge theory can 

%These results suggest that the framework of Lorentz gauge theory might provide the more robust foundation for a pregeometric unification of fundamental physics. 

%Consider the problem solved!
%In view of these features, 
%One of the very \1 steps, and an utterly necessary one, is the consistent formulation of the action principle. Already in classical considerations, one could maintain that the action principle should uniquely determine not only the dynamics (EoMs) but also the observables (charges). I
%Remarkably, the complete system of spacetime, matter and gauge fields is consistent, all the charges have a clear physical interpretation, involving no ``improvements'' or ambiguities. 

\begin{acknowledgments}
TSK is grateful to Prof. F. W. Hehl for useful remarks on history.  This work was supported by the Estonian Research Council grant PRG356 ``Gauge Gravity'' and the Centre of Excellence TK202 ``Foundations of the Universe''. TZ and LV acknowledge support from the project No.~2021/43/P/ST2/02141 co-funded by the Polish National Science Centre and the European Union Framework Programme for Research and Innovation Horizon 2020 under the Marie Sk\l{}odowska-Curie grant agreement No. 94533.
\end{acknowledgments}

\bibliography{CPSrefs}

\end{document}